\documentclass[paper, leqno]{ieice}
\usepackage{verbatim}

\usepackage{newtxtext}
\usepackage[varg]{newtxmath}
\usepackage{url}
\usepackage{booktabs}
\usepackage{listings}
\usepackage{caption}
\usepackage{amsmath}
\usepackage{float}
\DeclareCaptionFont{small}{\small}
\usepackage{graphicx}
\graphicspath{{./pdf/}{./jpeg/}{./png/}}
\DeclareGraphicsExtensions{.pdf,.jpeg,.png}
\usepackage{color}
\usepackage{flushend}
\usepackage{fancyvrb}

\definecolor{myred}{RGB}{135, 5, 5}

\field{}
\title{Real-time System Modeling and Verification through \\ Labeled Transition System Analyser (LTSA)}
\authorlist{%
 \authorentry[yylonly@gmail.com]{Yilong Yang}{m}{um}\MembershipNumber{1586718}
 \authorentry[xsl@umac.mo]{Xiaoshan Li}{n}{um}
 \authorentry[quanzu@umac.mo]{Quan Zu}{n}{um}
}
\affiliate[um]{The author is with the department of computer and information science, faculty of science and technology, university of Macau, Macau}

\DeclareCaptionFont{white}{\color{white}}
\DeclareCaptionFormat{listing}{\colorbox{gray}{\footnotesize\parbox{0.48\textwidth}{#1#2#3}}}
\definecolor{dkgreen}{rgb}{0,0.6,0}
\definecolor{gray}{rgb}{0.5,0.5,0.5}
\definecolor{mauve}{rgb}{0.58,0,0.82}

\lstdefinestyle{Java}{
  frame=b,
  language=Java,
  aboveskip=3mm,
  belowskip=3mm,
  showstringspaces=false,
  columns=flexible,
  basicstyle={\scriptsize\ttfamily},
  numbers=left,
  numberstyle=\tiny\color{gray},
  keywordstyle=\color{blue},
  commentstyle=\color{dkgreen},
  stringstyle=\color{mauve},
  breaklines=true,
  breakatwhitespace=true,
  tabsize=3
}

\definecolor{level2}{RGB}{180,30,80}
\definecolor{level3}{RGB}{36,148,227}
\definecolor{level4}{RGB}{120,120,120}
\definecolor{level5}{RGB}{150,150,150}

\lstdefinestyle{FSP}{
keywordstyle=\color{},
keywordstyle=[2]\color{level2}\bfseries,
keywordstyle=[3]\color{level3}\mdseries,
keywordstyle=[4]\color{level4},
keywordstyle=[5]\color{level5},
basicstyle={\tiny\ttfamily},
numberstyle=\tiny\color{gray},
frame=b,
columns=fullflexible,
showstringspaces=false,
morekeywords={},
keywords=[2]{progress, property, when, range, const},
keywords=[3]{STEAMBOILER, STEAMBOILERUN, WATERSENSOR, WATERSENSORRUN, STEAMSENSOR, STEAMSENSORRUN, PUMPSENSOR, PUMPSENSORRUN, PUMPCONTROLLER, PUMPCONTROLLERUN, SYSCONTROLRUN, CONTROLSYSTEM, TIMER},
keywords=[4]{in,out},
keywords=[5]{var}
}

\begin{document}
\maketitle
\begin{summary}
With the advancement of software engineering in recent years, the model checking techniques are widely applied in various areas to do the verification for the system model. However, it is difficult to apply the model checking to verify requirements due to lacking the details of the design. Unlike other model checking tools, LTSA provides the structure diagram, which can bridge the gap between the requirements and the design. In this paper, we demonstrate the abilities of LTSA shipped with the classic case study of the steam boiler system. The structure diagram of LTSA can specify the interactions between the controller and the steam boiler, which can be derived from UML requirements model such as system sequence diagram of the steam boiler system. The start-up design model of LTSA can be generated from the structure diagram. Furthermore, we provide a variation law of the steam rate to avoid the issue of state space explosion and show how explicitly and implicitly model the time that reflects the difference between system modeling and the physical world. Finally, the derived model is verified against the required properties. Our work demonstrates the potential power of integrating UML with  model checking tools in requirement elicitation, system design, and verification.


\end{summary}

\section{Introduction}

Real-time computing \cite{shin1994real} describes hardware and software systems depending not only on the logical correctness  of computation but also on the time constraints. The critical real-time systems must be dependable because any failure of the system could cause an economic disaster or even loss of human lives. The safety properties of such systems must be verified in the design stage as well as the testing stage of development before any deployment. Formal methods \cite{oda2017formal}\cite{saeeiab2000method} provide promising approaches to verify the safety properties of system. The papers \cite{heitmeyer1993benchmark}\cite{pajic2012verification}\cite{calabrese2011real} provide the case studies of general railroad crossing problem, pacemaker, and cell phone, to illustrate how to verify the safety properties of the real-time systems. Unlike mathematical approach to prove safety properties, model checking \cite{Ben-Ari:2010:PMC:1721933.1721950} is a promising approach, which can automatically verify the safety properties by exploring the state space of the target system.  For example, UPPAAL \cite{Behrmann2004} is one of the successful model checking tools widely applied to various areas from communication protocols to multimedia applications, in particular to the real-time system. When using UPPAAL for verifying the target system,  the components of the system are abstractly modeled as finite state machines based on the theory of timed automata \cite{Alur1990}. The advantages are more intuitive and easier understanding while modeling the real-time system as state machines. However, representing state machines graphically severely limits the complexity of problems that can be addressed. Consequently, textual process calculuses such as Calculus of Communicating Systems (CCS) \cite{milner1989communication} and Communicating Sequential Processes (CSP) \cite{Hoare2002} as the compensations of the finite state machine are usually used for modeling complex systems. 

In this paper, we use Labeled Transition System Analyser (LTSA) to model real-time systems and verify safety properties \cite{magee1999state}. Because LTSA contains many charming features: 1) LTSA provides both graphic state machine and textual process calculus, Finite State Processes (FSP), to model systems. 2) A textual process model can be automatically transformed to the corresponding graphic state machine. 3) LTSA can describe system architectures by structure diagrams (simplified form of the graphical representation of Darwin \cite{darwin1994}), which can describe the high-level model design and the static structure. 4) The skeleton of process model can be automatically generated from system architectures for the next round fine-grained design \cite{darwinToFSP}. One interesting work \cite{miyazaki2012synthesis} proposes a method to detect refinement errors in UML sequence diagrams using LTSA. That demonstrate the power the LTSA for UML model checking.

Although LTSA has many advantages for modeling system, there are few examples of real-time systems modeling through LTSA. The steam boiler \cite{abrial1996steam}\cite{abrial1996formal} is a classic case study for real-time system modeling in many studies.  The steam boiler is the minimal real-time system that contains all the essential parts of a real-time system: a controller and a controlled object with the sensors and the actuators.  The controller can periodically sample the state of the controlled object by the sensors, then strategically change the state of the controlled object through the actuators to guarantee the safety of the whole system. All the components of the real-time system are synchronized through time. In the physical world, time is implicitly contained in the physical phenomenon; e.g., the current temperature of the water was determined by the heating power and the heating time. Therefore, when modeling the real-time system, the time should be considered as inside of physical law of the controller object. However, for abstraction of the physical world and easy understanding, many studies explicitly model the time as a component of the system named timer, and then use that timer component to synchronize with other components of the system.


\vspace{.1cm}
\noindent\textbf{Related Work:} As a classic real-time system, the steam boiler has been widely studied. The paper \cite{woodcock2001steam} presents a formalization of the steam boiler problem using Circus, which is a unified theory of Z and CSP. It utilizes the strength of Z notation to describe the specifications of the system and their refinement, the strength of CSP to describe and reason about concurrency. Using Timed Automata to solve the steam boiler problem is mentioned in \cite{leeb1996proving}, which describes time constraints in the model with clocks. The guards may enable or disable transitions according to clock values. Mean Value Calculus models the steam boiler in \cite{xiaoshan1996specifying}, which could be used to specify and reason about time and logical constraint of the real-time system. Other related case studies for modeling and verifying real-time system are: Spin \cite{loffler1997creating} Signal-coq \cite{kerboeuf2000specification}, Action System \cite{butler1996action}, PLUSS \cite{gaudel1996formal},FOCUS \cite{broy1998streams}, LOTOS \cite{carreira2000automatically} and Hybrid Relation Calculus \cite{he2013hybrid}. Although the steam boiler problem has been elaborately studied, almost all case studies above have the following defects: 1) they explicitly model time as a timer component, that will make a gap between the real-time system design and the physical real-time system. 2) they do not provide a variation law for the steam rate, that will lead to the issue of state space explosion. 3) they do not contain any diagram to describe the system architecture, that will make it hard to directly elicit the components from target problem.

\vspace{.3cm}
\noindent\textbf{Contributions:} Although some tools support implicit time modeling inside of the system, to our best knowledge, there is no related work to illustrate the differences between the implicit timer and explicit timer, and the relationship between the physical world timer and timer in the real-time modeling. the Contributions of our paper are:

\vspace{.3cm}
\noindent 1) Demonstrating LTSA modeling and verifying abilities for real-time system shipped by a classical steam boiler problem.

\vspace{.1cm}
\noindent 2) Demonstrating how to specify structure diagram from UML requirement model, then generate the sketch of design model from structure diagram by Darwin.

\vspace{.1cm}
\noindent 3) Providing a variation law for the steam rate to avoid the issue of state space explosion.

\vspace{.1cm}
\noindent 4) Discussing how to model the explicit and implicit timer in the real-time system, and the relationship of time between the system modeling and physical world.

\vspace{.3cm}
\noindent The remainder of this paper is organized as follows: Section 2 is preliminary of FSP specification and overviews the steam boiler problem. Section 3 shows the interaction requirements and the derived structure diagram. Section 4 presents how to model the steam boiler in LTSA, especially for the time modeling. And then Section 5 shows the LTSA verification and simulation. Finally, section 6 concludes this paper and puts forward the future work.


\section{Preliminary}
LTSA adopts FSP as textual model to describe the system. To make this paper self-contained, we present the specification of FSP and the brief introduction of steam boiler problem in this section. More details of FSP could be found in the textbook \cite{magee1999state}.

\subsection{The Specification of FSP}
Finite State Process (FSP) is CSP syntax-liked formal language for modeling concurrency system \cite{magee1999state}, it uses concept of \textit{Primitive Process} to define the component of the system which may contain the sequences of \textit{actions}. Component composition could be defined as \textit{Composited Processes} in which concurrent executions of actions could be synchronied or interleaved. The requirements of system could be captured as the \textit{Properties} of FSP. Once both properties and processes of the system are defined, LTSA can check the satisfiability of properties for particular system. The brief summary of FSP specification is provided as follows:

\vspace{.3cm}
\noindent \textbf{Primitive Process} A primitive process is the execution of a sequential program. The state of primitive process is transformed by executing actions. We use primitive process to define the component of the system. Like any programming languages, primitive process may contain choice expression, condition, 

\begin{itemize}
\item Action Prefix \verb|->|:  (a  \verb|->| P) describes a process which engages in the action a and then behaves as described by P.
\item Choice \texttt{|}: (a \texttt{->} P \texttt{|} b \texttt{->} Q) describes a process which initially engages in either of the actions a or b. After the first action has been performed, the subsequent behaviour is described by P if the first event was a, or by Q if the first event was b.
\item STOP: It is sometimes (if rarely) necessary to specify a primitive process which terminates. Consequently, a local process STOP is predefined which engages in no further actions.
\item Alphabet Extension \verb|+{}|:  Each primitive process has an alphabet consisting of the actions it may take part in. A process may only engage in the actions contained in its alphabet although the converse does not hold. It is sometimes useful to extend the alphabet of a process with actions that it does not engage in and consequently actions that are not used in its definition. This may be done to prevent another process executing the action.
\item Indexing: Both local process names and action names may be indexed.
%
 Both local processes and actions may have more than one index as illustrated by this example (for actions). A process which outputs the sum of two integers (in the range 0..N).
\item Conditional: A conditional takes the form: if \textit{expr} then \textit{local\_process} else \textit{local\_process}.
\item Guards: A guarded transition takes the form (when B a \verb|->| P) which means that the action a is eligible when the guard B is true, otherwise a cannot be chosen for execution.
\end{itemize}

\vspace{.3cm}
\noindent \textbf{Composited Processes} Parallel Composition $\parallel$: (P $\parallel$ Q) expresses the parallel composition of the processes P and Q. It allows all the possible interleavings of the actions of the two processes. These shared actions synchronise the execution of the two processes. If the processes contain no shared actions then the composite state machine will describe all interleaving.

\vspace{.3cm}
\noindent \textbf{Safety Properties} A safety property asserts that nothing bad happens. Informally, a property process specifies a set of acceptable behaviours for the system it is composed with. A system S will satisfy a property P if S can only generate sequences of actions (traces) which when restricted to the alphabet of P, are acceptable to P.

\vspace{.3cm}
\noindent \textbf{Progress Properties} A progress property asserts that it is always the case that an action is eventually executed. We will define progress to check the steam boiler is still work or not.

\subsection{Steam Boiler Problem}
The steam boiler problem \cite{abrial1996steam}\cite{abrial1996formal} is one of the typical problems in software engineering, which divides the system into physical and control parts. For the physical part, it has a physical steam boiler, three water/steam/pump sensors and pumps. For the control part, it is a controller which could get the value from the sensors, make a decision, and send orders of switching on/off to the pumps. The communication between physical and control parts are messages.

\begin{table}[!htb]
  \centering
  \footnotesize
   \caption{The Summary of Constants and Variables}\label{SumaryofCV}
      \begin{tabular}{ccc}
      \toprule
      Type & Unit & Comment  \\
       \midrule
      $\mathit{Interval}$ & $\mathbb{N}$ & Sample Period/Delay of PumpOn \\
      \midrule
     \midrule
      Quantity of Water &   &    \\
      \midrule
      $C$ & litre & Maximal Capacity \\
      $M_1$ & litre & Minimal Limit \\
      $M_2$ & litre & Maximal Limit \\
      $N_1$ & litre & Minimal Normal Limit \\
      $N_2$ & litre & Maximal Normal Limit\\
       \midrule
       \midrule
       Outcome of Steam &   &    \\
      \midrule
      $W$  & litre/sec & Maximal Rate \\
      $U_1$ & litre/sec/sec & Increase Rate \\
      $U_2$ & litre/sec/sec & Decrease Rate\\
      \midrule
       \midrule
      Pump Parameters  &   &  \\
      \midrule
      $P$ & litre & Capacity of each Pump \\
      \midrule
       \midrule
      Current Variables &  &    \\
      \midrule
      $q$  & litre & Quantity of Water \\
      $v$ & litre/sec & Steam Rate \\
      $p$ & litre/sec	& Throughput of pumps \\

      \bottomrule
    \end{tabular}
\end{table}

The summary of constants and variables are in Table \ref{SumaryofCV}, constants are as follows: $Interval$ is the sample cycle and delay of \textit{PumpOn} action, $C$ is the maximal capacity of the steam boiler.  $M_1$ is minimal limit of water quantity, $M_2$ is maximal limitation, $N_1$ is minimal limitation and $N_2$ is maximal limitation in normal mode, $W$ is maximal quantity of outcome steam, the increase rate of outcome steam are defined by $U_1$,  the decrease rate of outcome steam is $U_2$,  the number of pumps is 5, $P$ is capacity of each pump.

System constants must satisfy the in-equation:
\[
\begin{array}{c}
0<M_1<N_1<N_2<M_2<C \\
\end{array}
\]

Variable $q$ represents the quantity of water, $v$ is current outcome rate of steam, $p$ is the current throughput of pumps. Invariants of variables must be satisfied:
\[
\begin{array}{c}
0 \le q \le C \\
0 \le v \le W \\
0 \le p \le 5P
\end{array}
\]

\section{Requirement Analysis}
\begin{figure}[htb]
  \centering
  \includegraphics[width=0.40\textwidth]{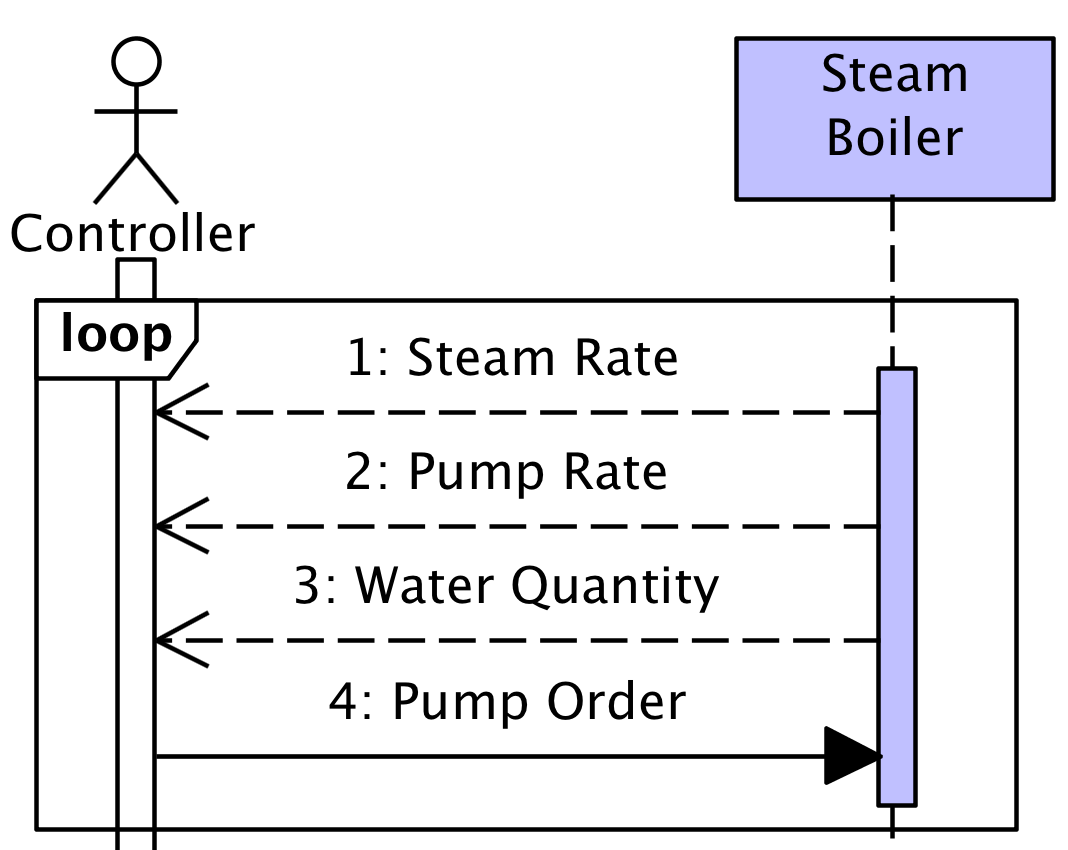}
  \caption{Interaction Requirements between Controller and Steam Boiler}\label{ssd}
\end{figure}

Before modeling the system, we need figure out the requirements of the steam boiler problem especially for the interactions between the control part \textit{Controller} and the physical part \textit{Steam Boiler}. System sequence diagram of UML in Fig.\ref{ssd} shows the steam boiler periodically sends the values of steam rate, pump rate, and water quantity to the controller, then the controller sends the pump order to the steam boiler according to the values of the measures $q, v, p$.

These interactions must make the water quantity of steam boiler keep between $N_1$ and $N_2$ in normal mode, and between $M_1$ and $M_2$ in the rescue mode when correct $q$ can not be obtained from the water sensor. Formally, the requirements (safety property) of the steam boiler system are described as follows:

\begin{equation}
\mathit{REQ_{NormalMode}} \; \widehat{=} \; N_1 \le q \le N_2
\label{normal}
\end{equation}
\begin{equation}
\mathit{REQ_{RescueMode}} \; \widehat{=} \; M_1 \le q \le M_2
\label{rescuereq}
\end{equation}


\vspace{.3cm}
\noindent \textbf{FSP Structure Diagram}
\vspace{.1cm}

\begin{figure}[htb]
  \centering
  \includegraphics[width=0.48\textwidth]{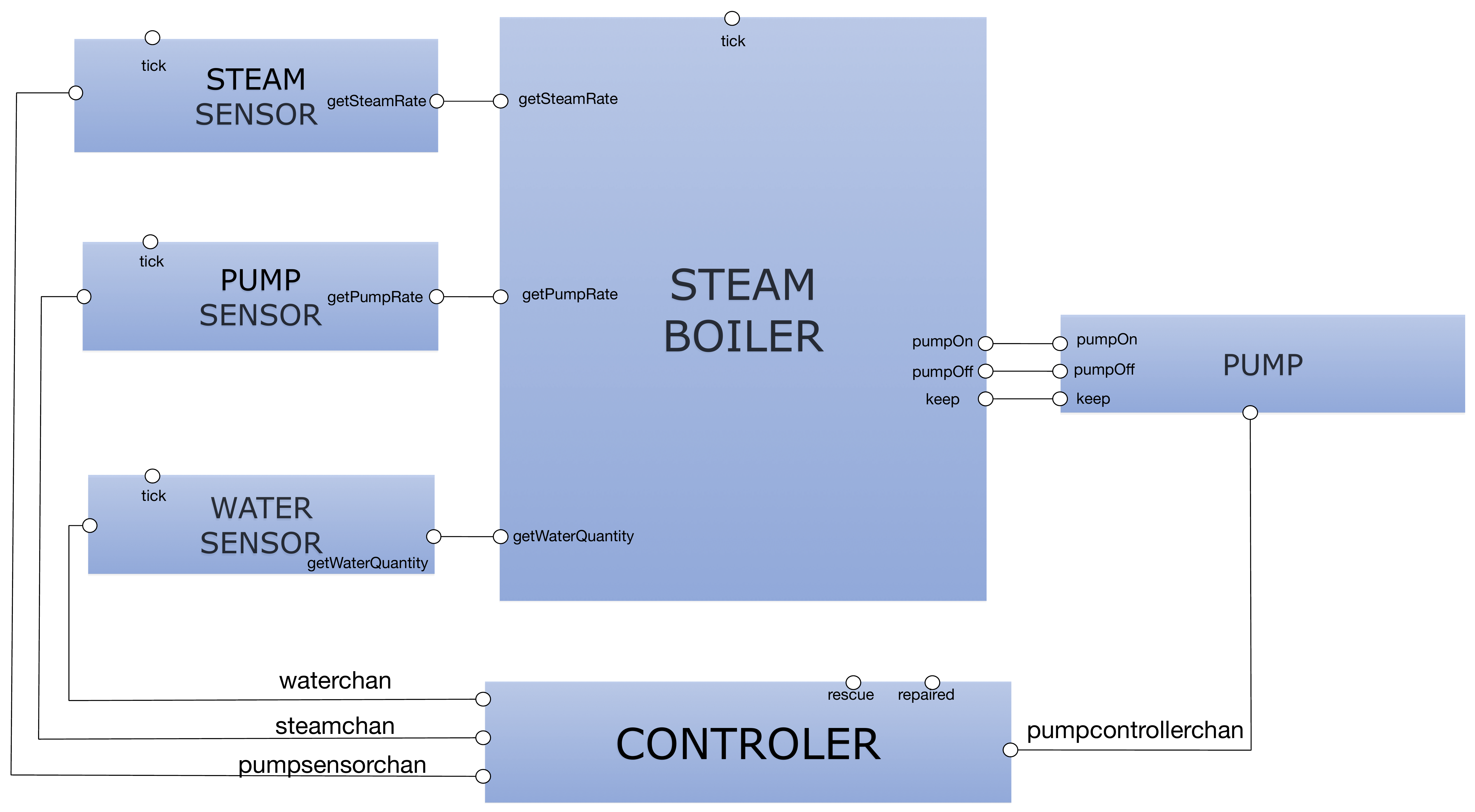}
  \caption{FSP Structure Diagram of the Steam Boiler}\label{StructureDiagram}
\end{figure}

\noindent From the system sequence diagram in Fig.\ref{ssd}, we can smoothly derive the FSP structure diagram in Fig.\ref{StructureDiagram}. This diagram shows all the components and their actions in the system. The component \textsf{STEAM BOILER} represents the physical steam boiler, which is assumed to keep boiling (boiling action) all the time. The state of \textsf{STEAM BOILER} includes the quantity of water, the quantity of steam and the throughput of the pumps, which are respectively denoted by the variables $p$, $v$, $q$. They are measured in a fixed sampling cycle via the sensors \textsf{STEAM SENSOR}, \textsf{PUMP SENSOR} and \textsf{WATER SENSOR} by the actions \textit{getPumpRate}, \textit{getSteamRate} and \textit{getWaterQuantity} respectively. The \textsf{STEAM BOILER}'s state is changed according to the state of \textsf{PUMP}, which is controlled by the actions \textit{pumpOn}, \textit{pumpOff} and \textit{keep}. In each sampling cycle, the control system \textsf{CONTROLLER} receives the measures of $p$, $v$, $q$, according to which \textsf{CONTROLLER} will decide the message (\textit{pumpOn} or \textit{pumpOff} or \textit{keep}) sent to \textsf{PUMP} through the channel \textit{pumpcontrollerchannel}. Each message triggers the corresponding action in \textsf{PUMP}.

\section{Model Design}
In the previous section, we present the requirements of steam boiler as well as the derived FSP structure diagram. In this section, we first automatically generate FSP component processes from the FSP structure diagram by Darwin \cite{darwin1994}, then design the details of the each component. From the FSP structure diagram, \textsf{SYSTEMDESIGN} is generated as:



\begin{Verbatim}[fontsize=\scriptsize]
||SYSTEMDESIGN = (STEAMBOILER || CONTROLSYSTEM || PUMPCONTROLLER || 
		WATERSENSOR || STEAMSENSOR || PUMPSENSOR || TIMER)
\end{Verbatim}

\noindent The sets of actions of each component are defined as:
\begin{Verbatim}[fontsize=\scriptsize]
set Timer = {tick}
set PumpSensor = {tick, getPumpRate, pumpsensorchan}
set SteamSensor = {tick, getSteamRate, steamchan}
set WaterSensor = {tick, getWaterQuantity, waterchan}
set Pump = {pumpOn, pumpOff, keep, pumpcontrollerchan.}
set SteamBoiler = {tick, getPumpRate, getSteamRate,
             getWaterQuantity, pumpOn, pumpOff, keep}
set Controller = {rescue, repaired, makedecision, waterchan,
	pumpsensorchan, steamchan, pumpcontrollerchan}	
\end{Verbatim}

\noindent The remaining subsections specify each component in details.

\subsection{Timer}

Assume there is a start time denoted by $t_0$. We use $t_i$, $i\in \mathbb{N}$, to denote the time point which is elapsed $i$ seconds since the start time. Therefore, the trace of time is represented as $\langle t_0, t_1, ... , t_i, ..., t_n \rangle$ in the system. For example, since the sample period is 5 seconds in this system, the next sampling point will be $t_{i+5} = t_i + 5$ if the previous sampling point is $t_i$. For $q$ and $v$, they would be changed in every second. For $p$, if sampling starts from time point $t_i$, sampling period is 5 seconds, it would only be changed in time points $\{t_{i+5j} \, | \, j \in \mathbb{N} \}$.

In order to model the time in the system, we use a component \textsf{TIMER} that synchronizes with \textsf{STEAM BOILER} and all the sensor components by the action \textit{tick}. \textsf{TIMER} in FSP form is: 
\begin{Verbatim}[fontsize=\footnotesize]
TIMER =  (tick -> TIMER)
\end{Verbatim}
where each \textit{tick} represents the pass of one second. 

If the sampling period is identical with the delay of pumping, which is the case in the steam boiler specification\cite{abrial1996steam}, we can model the system in a time-implicit way. That is, we don't need a \textsf{Timer} component to explicitly specify the time passing. We compare these two different modeling methods in the next subsection.

\subsection{Steam Boiler Component}
The specification\cite{abrial1996steam} specifies part of the behaviours in the steam boiler system. For instance, after switching on the pump, the water starts pouring into the boiler in 5 seconds. But some details are not given, including the variation law of the steam rate and the control strategy of the pump. We design a pumping control strategy in next subsection. In this subsection, we hypothesize a physical variation law for the steam rate and show the model of the steam boiler in FSP.

\subsubsection{Quantity of Water}
We use $q_i, v_i, p_i$ to respectively denote the steam rate, the quality of water, and the pumping rate at the $t_i$ time point. Then clearly we have the following equation:
\begin{equation}
q_{i+1} = q_i + (p_i - v_i) * \Delta t
   \label{qfunction}
\end{equation}
where $\Delta t = 1$. 

\subsubsection{Steam Rate}
\begin{figure}[htb]
  \centering
  \includegraphics[width=0.35\textwidth]{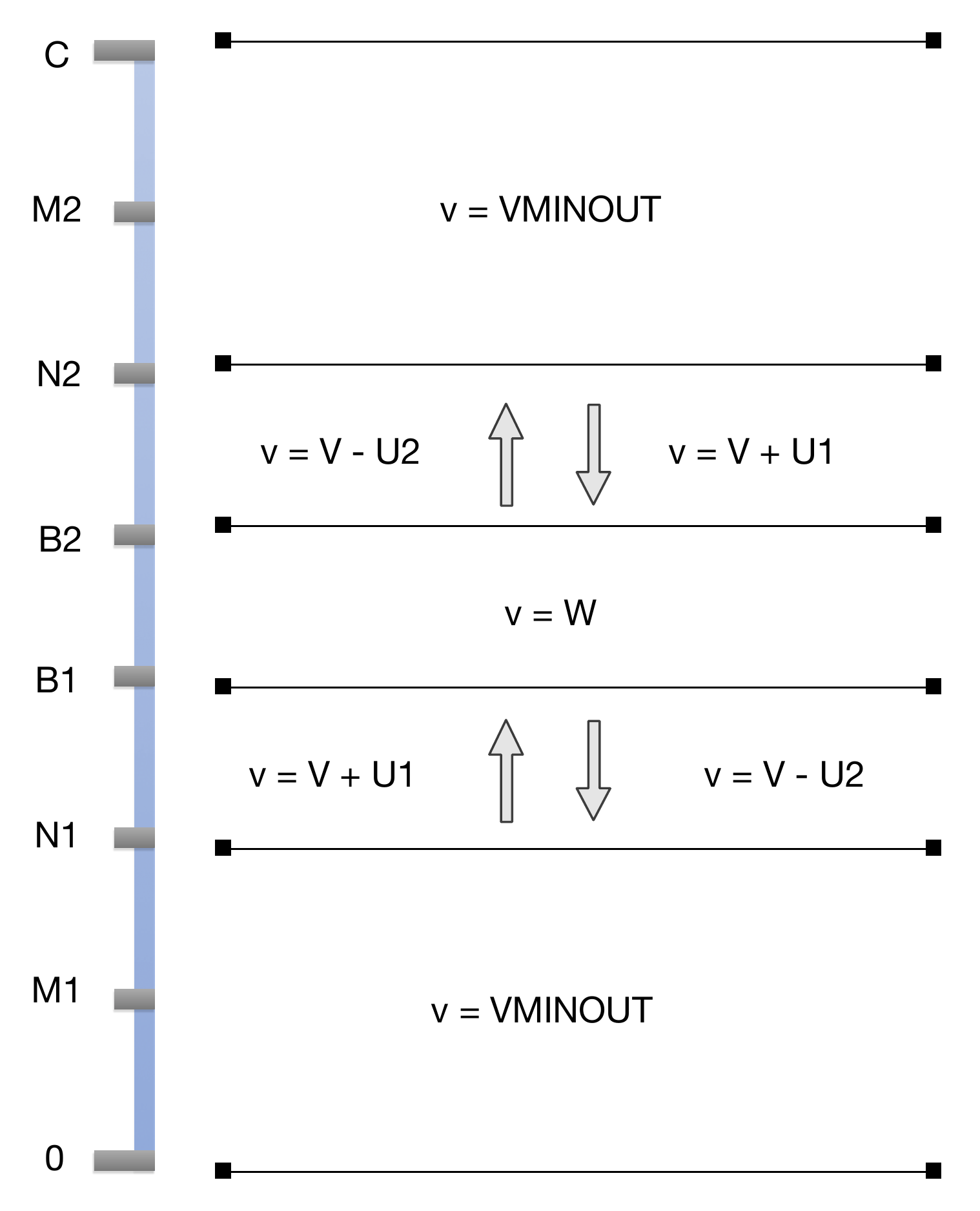}\\
  \caption{Steam Rate}\label{steamrate}
\end{figure}

The hypothesis of the law of the steam rate is based on the fact that the steam rate is influenced by the quantity of the water. Between $N_1$ and $N_2$, we add another two quantities of water level $B_1$ and $B_2$, the best minimal limit and the best maximal limit, which are used to construct the hypothesis. We use $\textsf{VMINOUT}$ to denote the minimal steam rate. It is required that $(N_1 < B_1 < B_2 < N_2)$ and  $0< \textsf{VMINOUT} < min\{U_1, U_2\}$ are hold. We use $v_i$ to denote the steam rate at the $t_i$ time point. The law of steam rate is depicted in Figure \ref{steamrate} and Function \ref{vfunction}.

If current water level is between $N_2$ and $C$ or between $N_1$ and 0, the steam rate is minimal, denoted by $\textsf{VMINOUT}$. If current water level is between $B_1$ and $B_2$, the rate is maximal, denoted by $W$. If current water level is between $N_1$ and $B_1$ and it is increasing ($p_i - v_i > 0$), or if current water level is between $B_2$ and $N_2$ and it is decreasing ($p_i - v_i < 0$), the steam rate will have an increment of $U_1$. If current water level is between $N_1$ and $B_1$ and it is decreasing ($p_i - v_i < 0$), or if current water level is between $B_2$ and $N_2$ and it is increasing ($p_i - v_i > 0$), the steam rate will have a decrement of $U_2$. 

 {\footnotesize
\begin{equation}
   v_{i+1} =
   \begin{cases}
   \textsf{VMINOUT} &\mbox{if } N_2 \le q_i \le C \mbox{ or }  0 \le q_i  \le N_1  \\
   v_i - U_2 & \mbox{if } B_2 \le q_i < N_2 \mbox{ and } p_i - v_i > 0  \\
   v_i - U_2 & \mbox{if } N_1 < q_i < B_1 \mbox{ and } p_i - v_i < 0 \\
   v_i + U_1  & \mbox{if } B_2 \le q_i < N_2  \mbox{ and } p_i - v_i < 0 \\
   v_i + U_1 & \mbox{if } N_1 < q_i < B_1 \mbox{ and } p_i - v_i > 0  \\
   W & \mbox{if } B_1 \le q_i \le B_2
   \end{cases}
   \label{vfunction}
\end{equation}  }

\subsubsection{Throughput of pumps}
It is specified that, after switching on, the pump needs 5 seconds to pour the water into the boiler. That is, assuming the action \textsf{pumpOn} is trigger at $t_i$, the pumping rate $p$ will have a increment of $P$ at $t_{i+5}$. To model this time delay, we use a sequence of 5 bits to represent the time points at which the pump rate will increase. Each bit in the sequence represents the action in the corresponding time point. For instance, $10000$ means the pump will increase in 5 seconds, and $00101$ means the pump will increase now and in 3 seconds. We assume that there is at most one pump opening in one second. Once receiving the action \textsf{pumpOn}, the first bit of the sequence will be set to 1. In each second, the sequence will move one position to the right, representing one second elapses.

\subsubsection{Steam Boiler in FSP}
\textsf{STEAMBOILER} has two subcomponents \textsf{STEAMBOILERUN} and \textsf{PUMPDELAY}. The sensors get the measures from the steam boiler by the actions \textit{getPumpRate}, \textit{getSteamRate} and \textit{getWaterQuantity}. The steam boiler communicates with the pump by the actions \textit{pumpOn}, \textit{pumpOff} and \textit{keep}. If \textit{pumpOff}, the pump rate decreases immediately. If \textit{pumpOn}, $t$ is increased by $16$, which is $10000$ in binary. The steam boiler synchronizes with \textsf{Timer} by the action \textit{tick}. In each second, the quantity of the water is changing according to Formula \ref{qfunction} and the steam rate changes according to the law. The pumping rate will be checked in \textsf{PUMPDELAY}. It firstly checks whether the last bit is 1, representing current increment. Then it moves the sequence one position to the right by the division by 2. The following is part of the model.

\begin{Verbatim}[fontsize=\scriptsize, numbers=left, numbersep=5pt]
STEAMBOILER = (start->STEAMBOILERUN[INITQ][W][PUMPQ][0]),
    STEAMBOILERUN[q:Q][v:V][p:PUMPQ][t:PUMPMAXDELAY] = (
      getWaterQuantity[q] -> getSteamRate[v] 
        -> getPumpRate[p] -> STEAMBOILERUN[q][v][p][t]
    | pumpOn -> STEAMBOILERUN[q][v][p][16+t] 			
    | pumpOff -> STEAMBOILERUN[q][v][p-PQ][t]
    | keep -> STEAMBOILERUN[q][v][p][t]
    | tick -> ( 
       when (q >= N2 )
        boiling -> PUMPDELAY[q+(p-v)][VMINOUT][p][t]
       | when (BEST2 < q && q < N2 && (p-v) < 0 && (v+UP) < W)
        boilingBEST2toN21[q][v][p] -> PUMPDELAY[q+(p-v)][v+UP][p][t]
        ...... ) ),
    PUMPDELAY[q:Q][v:V][p:P][t:PUMPMAXDELAY] = (
      when (t % 2 == 0) pumping -> STEAMBOILERUN[q][v][p][t/2]					
    | when (t % 2 != 0) pumping -> STEAMBOILERUN[q][v][p][(t-1)/2]).
\end{Verbatim}

\subsubsection{Throughput of pumps in implicit time}
In the case of the sampling period is identical with the pumping delay time, we can use the implicit way to model the system. For the function of $p$, when the action \textit{pumpOn} is triggered at $t_i$, the fixed delay is required before $p$ is changed. Therefore, $p$ is unchanged from the time point $t_{i}$ to $t_{i+4}$. The value of $p_{i+5}$ is determined by the following factors: 1) $p_i$, 2) whether the last pump order $lastpo \in \mathit{\{True, False\}}$ is \textit{pumpOn} or not, 3) the previous order $actionp \in \mathit{\{pumpOn, pumpOff, keep\}}$, and 4) the current order $actionc \in \mathit{\{pumpOn, pumpOff, keep\}}$ of the pump.

{\footnotesize
\begin{equation}
   p_{i+5} =
   \begin{cases}
   p_i + P  &\mbox{if } \mathit{lastpo} = \mathit{True}  \mbox{ and }  \mathit{actionc} = \mathit{pumpOn}  \\
   p_i + P		 &\mbox{if } \mathit{lastpo} = \mathit{True}  \mbox{ and }  \mathit{actionc} = \mathit{keep}  \\
   p_i  &\mbox{if } \mathit{lastpo} = \mathit{True}  \mbox{ and } \mathit{actionc} = \mathit{pumpOff}  \\
   p_i 	  &\mbox{if } \mathit{lastpo} = \mathit{False}  \mbox{ and }  \mathit{actionc} = \mathit{pumpOn}  \\
   p_i 	  &\mbox{if } \mathit{lastpo} = \mathit{False}  \mbox{ and }  \mathit{actionc} = \mathit{keep}  
   \end{cases} 
   \label{pfunction}
\end{equation} 
\begin{equation}
p_{i+1} = 
	\begin{cases}
		p_i - P  & \mbox{   if } \mathit{lastpo} = \mathit{False}  \mbox{ and }  \mathit{actionc} = \mathit{pumpOff}  \\
		p_{i}  & \mbox{Otherwise} \\
	 \end{cases} 	
\end{equation}
\begin{equation}
   \mathit{lastpo} =
   \begin{cases}
      	\mathit{True} &\mbox{if } \mathit{actionp} = \mathit{pumpOn}  \\
   	\mathit{False} &\mbox{if } \mathit{actionp} \in \mathit{\{keep, pumpOff\}}  \\
   \end{cases}
   \label{lofunction}
\end{equation}
}

If the last order is $\mathit{pumpOn}$ and current order is $\mathit{pumpOn}$, $p_{i+5}$ is $p_i$ plus $P$. If the last order is $\mathit{pumpOn}$ and current order is $\mathit{keep}$, $p_{i+5}$ is $p_i$  plus $P$. If the last order is $\mathit{pumpOn}$, and current order is $\mathit{pumpOff}$, $p_{i+5}$ will not change.  It is same in the case that if the previous order is not $\mathit{pumpOn}$ and current order is $\mathit{pumpOn}$ or $\mathit{keep}$.  $p_{k+1}$ is $p_k$ minus $P$, if the last order is not $\mathit{pumpOn}$ and current order is $\mathit{keep}$. The reader may refer to the details of the implicit model in our website\footnote{\url{http://lab.mydreamy.net}}.

\subsection{Sensor Components}
Sensor devices measure the state of the steam-boiler and the value of the measures are transmitted to the control system through network or cable. The measure processes are modelled as the synchronizations between \textsf{STEAM BOILER} and the sensor components, and the transmissions are modeled as the synchronizations between \textsf{CONTROLLER} and the sensor components. Sensor components FSP processes are described as below:
\begin{Verbatim}[fontsize=\scriptsize]
WATERSENSOR = ( getWaterQuantity[q:Q] -> waterchan.send[q] ->
        tick -> tick -> tick -> tick -> tick -> WATERSENSOR).
STEAMSENSOR = ( getSteamRate[v:V] -> steamchan.send[v] ->
        tick -> tick -> tick -> tick -> tick -> STEAMSENSOR).
PUMPSENSOR = ( getPumpRate[p:5*P] -> pumpsensorchan.send[p] -> 
        tick -> tick -> tick -> tick -> tick -> PUMPSENSOR).
\end{Verbatim}

\subsection{Pump Component}
\begin{figure}[htb]
  \centering
  \includegraphics[width=0.48\textwidth]{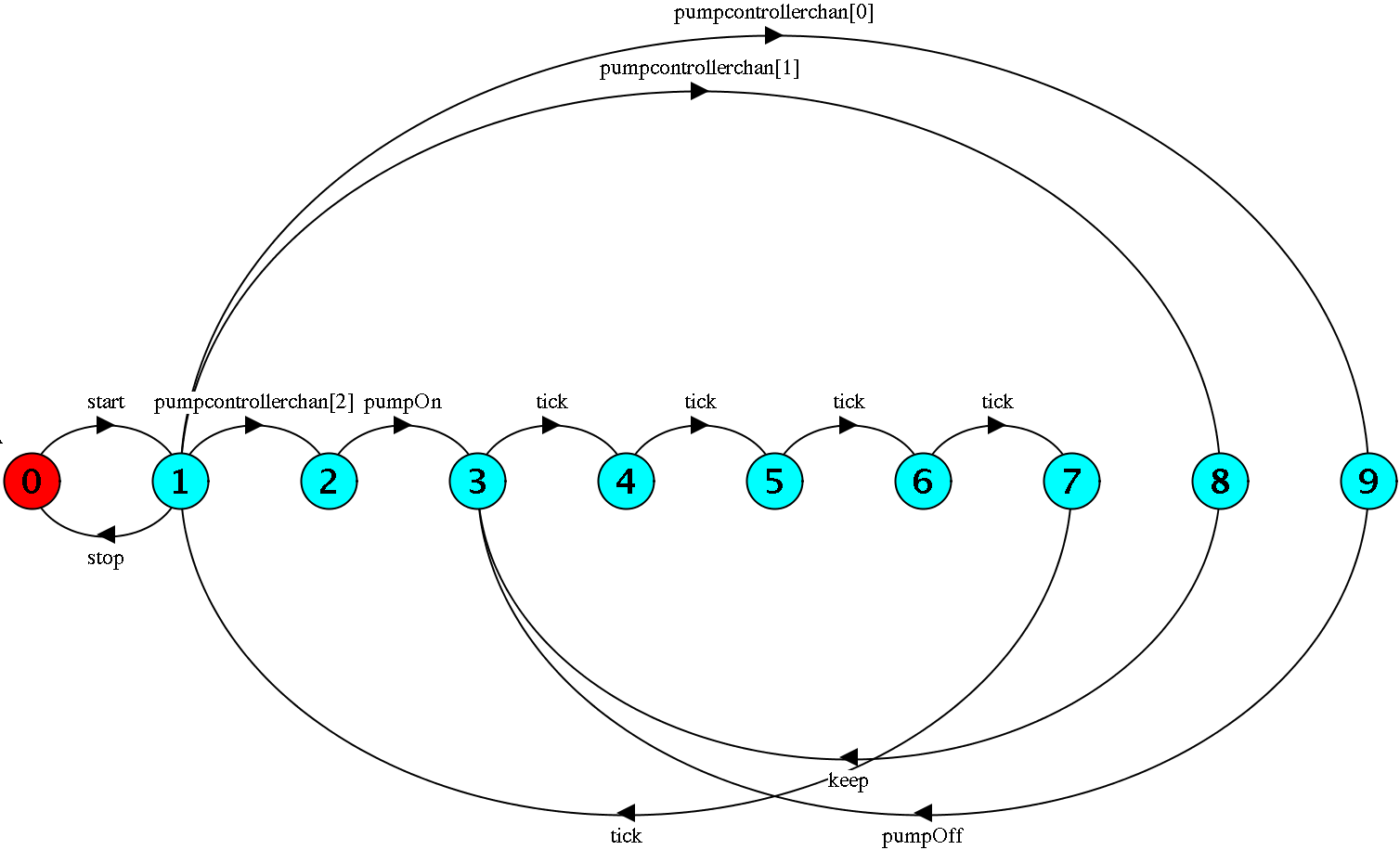}\\
  \caption{Pump Component in LTSA}\label{pumplts}
\end{figure}

The pump controller component is defined as: 
\begin{Verbatim}[fontsize=\scriptsize]
PUMPCONTROLLER = (pumpcontrollerchan.receive[o:PUMPORDER] ->
   (when (o == ON)   pumpOn ->  PUMPTICK |
    when (o == OFF)  pumpOff -> PUMPTICK |
    when (o == KEEP) keep -> PUMPTICK)),
PUMPTICK = (tick -> tick -> tick -> tick -> tick -> PUMPCONTROLLER).
\end{Verbatim}
Pump controller would do the action corresponding to the order $o \in \{\mathit{ON, KEEP, OFF}\}$ received from pump controller channel. Furthermore, the LTS form of pump component in Figure \ref{pumplts}.

\subsection{Controller Component}
The controller is the key component in the steam boiler system, and an appropriate strategy should make the quality of water $q$ change between a specific range. After receiving the sensor measures $q$, $v$ and $p$, the controller component will do the action \textit{makedecision} to generate an order $o$, then do action \textit{pumpcontrollerchan.send[o]} to send order $o$ to pump controller channel, function $o$ is defined as:

{\footnotesize
\begin{equation}
   o =
   \begin{cases}
      	\textsf{ON} & \mbox{if }  q < B_1+\textsf{FTRD} \mbox{ and } p \le 3P\\
	\textsf{ON} & \mbox{if }  q < B_1+\textsf{FTRD} \mbox{ and } p = 4P \mbox{ and } lastpo = \textsf{False}\\
	\textsf{KEEP} & \mbox{if }  q < B_1+\textsf{FTRD} \mbox{ and }  p = 5P \\
	\textsf{KEEP} & \mbox{if }  q < B_1+\textsf{FTRD} \mbox{ and }  p \le 4P  \mbox{ and } lastpo = \textsf{True}\\
	\textsf{KEEP} & \mbox{if }  B_1+\textsf{FTRD} \le q \le B_2-\textsf{FTRU}  \\
	\textsf{KEEP} & \mbox{if }  q > B_2+\textsf{FTRU}  \mbox{ and } v \ge 0 \mbox{ and } p = 0 \\
   	\textsf{OFF} &\mbox{if }  q > B_2+\textsf{FTRU}   \mbox{ and } (p-v) \ge 0 \mbox{ and } p > 0
   \end{cases}
   \label{makedecision}
\end{equation}}

Because of the delay of pumping, thresholds of \textsf{FTRD} and \textsf{FTRU} are introduced ($0 \le \textsf{FTRD} \le B_2 - B_1$, $0 \le \textsf{FTRU} \le B_2 - B_1$). If the water level is above $B_2$ plus \textsf{FTRU}, and the water level is not decreasing and the pump rate is greater than zero, the decision \textsf{OFF} is made. If the water level is under $B_1+\textsf{FTRD}$, and either throughout of pumps is less than $3P$ or throughout of pumps is $4P$ without the previous order \textit{pumpON}, the decision is \textsf{ON}. The decision is \textsf{KEEP}, if water level is between that two boundaries, or water level is above $B_2$ plus \textsf{FTRU} besides steam rate is great than zero, throughput of pumps is zero, or water level is under $B_1$ plus \textsf{FTRD} besides throughput of pumps is maximal or $4P$ and previous order is \textit{pumpOn}. Controller component of FSP is:
\begin{Verbatim}[fontsize=\scriptsize, numbers=left, numbersep=7pt]
CONTROLSYSTEM = (init -> SYSCONTROLRUN[OFF]),
SYSCONTROLRUN[po:PUMPORDER] = (waterchan.receive[q:Q] ->
     steamchan.receive[v:V] ->  pumpsensorchan.receive[p:5*P] ->
     makedecision ->
---------------------- q < BEST1+FTRD -------------------------
   ( when (q < BEST1+FTRD && p <= 3*P))
	pumpcontrollerchan.send[ON] -> CONTROLTICK[True][q]
   | when (q < BEST1+FTRD && p == 4*P && lastpo == False)
	pumpcontrollerchan.send[ON] -> CONTROLTICK[True][q]
   | when (q < BEST1+FTRD && p <= 4*P && lastpo == True)
	pumpcontrollerchan.send[KEEP] -> CONTROLTICK[False][q] 
   | when (q < BEST1+FTRD && p == 5*P)
	pumpcontrollerchan.send[KEEP] -> CONTROLTICK[False][q] 
--------------------- q > BEST2-FTRU --------------------------
   | when (q > BEST2-FTRU && (p-v) >= 0 && p > 0)
	pumpcontrollerchan.send[OFF] -> CONTROLTICK[False][q] 
   | when (q > BEST2-FTRU && v >= 0 && p == 0)	
	pumpcontrollerchan.send[KEEP] -> CONTROLTICK[False][q] 
----------------BEST1+FTRD <= q && q <= BEST2-FTRU--------------
   | when (BEST1+FTRD <= q && q <= BEST2-FTRU)
	pumpcontrollerchan.send[KEEP] -> CONTROLTICK[False][q])),

CONTROLTICK[o:PUMPORDER][q:Q] = (tick -> tick -> tick -> tick 
                	-> tick -> SYSCONTROLRUN[o][q]).

\end{Verbatim}
The boundary situations are in Line 17-18 and Line 25-26, and when the \textit{pumpOn} decision is made, the control system of FSP takes previous order into account in Line 8-18.

\section{Model Verification and Simulation}
This section verifies the FSP model against the safety properties and the progress properties in LTSA. Safety property checking guarantees that there is no deadlock in the system and the water quantity level keeps in the specified ranges. Progress property checking guarantees that there is no local loop in the system state.

\subsection{Requirement specified by FSP}
\noindent\textbf{Safety Property} For safety checking, basic and normal properties according to Requirement \ref{normal} and \ref{rescuereq} are defined for a different mode. FSP does not provide a mechanism for describing invariant directly. Hence, actions with parameter are used to describe invariants. Basic property is described as \textsf{getWaterQuantity[q:M1..M2]}, which requires that water quantity $q$ must be maintained between $M_1$ and $M_2$, normal property as \textsf{getWaterQuantity[q:N1..N2]} requires that water quantity $q$ must be retained between $N_1$ and $N_2$:
\begin{Verbatim}[fontsize=\scriptsize]
property BASIC = (getWaterQuantity[q:M1..M2] -> BASIC)
	+ {getWaterQuantity[0..M1-1], getWaterQuantity[M2+1..C]}.
property NORMAL = (getWaterQuantity[q:N1..N2] -> NORMAL)
	+ {getWaterQuantity[0..N1-1], getWaterQuantity[N2+1..C]}.
property OPTIMIZATION = (getWaterQuantity[q:BEST1..BEST2] 
	-> OPTIMIZATION) + {getWaterQuantity[0..BEST1-1], 
	getWaterQuantity[BEST2+1..C]}.		
\end{Verbatim}
Notes: Optimization property is defined for checking our model works. In the normal mode, our model must not violate basic and normal properties, but optimization property.  In the rescue mode, our model must not violate basic property within constraint the time, but normal and optimization properties.

\vspace{.3cm}
\noindent\textbf{Progress Property} In our case, the steam boiler system is in progress, when actions of boiling and boiling out of the steam boiler, \emph{makedecision} of the controller, get water quantity of water sensor, get the steam rate of the steam sensor, \emph{pumpOn} and \emph{pumpOff} of pumps are eventually executed. Those properties are defined:
\begin{Verbatim}[fontsize=\scriptsize]
progress WaterSensorWorking = {getWaterQuantity[q:Q]}
progress SteamSensorWorking = {getSteamRate[v:V]}
progress PumpSensorWorking = {getPumpRate[v:V]}
progress PumpControllerWorking = {pumpOn, pumpOff, keep}
progress CSWorking = {makedecision, makerescuedecision, rescue,
		      repaired}
progress STEAMBOILERWorking = {boiling[q:Q][v:V][p:P],
                               boilingout[q:Q][v:V][p:P]}
\end{Verbatim}

\noindent The defined properties must be compositied with \textsf{SYSTEMDESIGN} for verification:
\begin{Verbatim}[fontsize=\scriptsize]
||BASICSYSTEM = (SYSTEMDESIGN || BASIC).
||NORMALSYSTEM = (SYSTEMDESIGN || NORMAL).
||OPTIMIZATIONSYSTEM = (SYSTEMDESIGN || OPTIMIZATION).
\end{Verbatim}
After compositied, the processes \textsf{BASICSYSTEM, NORMALSYSTEM, and OPTIMIZATIONSYSTEM} will be verified in LTSA tools.

\subsection{LTSA Verification}
\begin{figure}[!htb]
  \centering
  \includegraphics[width=0.45\textwidth]{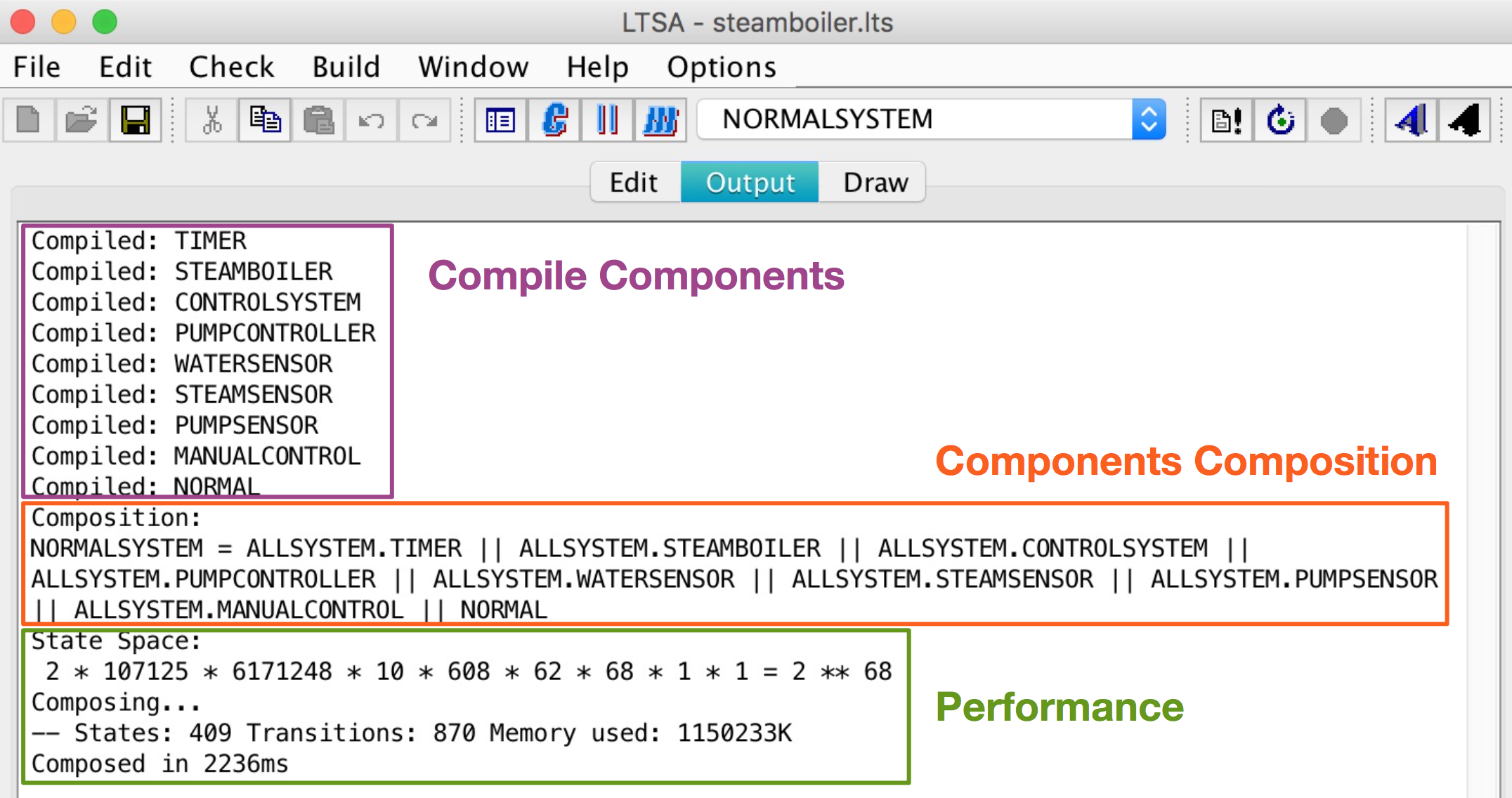}
  \caption{Verification for Normal Property}\label{normalp}
\end{figure}

In this section, FSP model of the steam boiler is loaded into LTSA\footnote{\url{http://www.doc.ic.ac.uk/ltsa/}}  (Labelled Transition System Analyser) for verification. We use the standalone version 3.0, the eclipse plugin could use as the same way as well. \textsf{NORMAL} property  $N_1 \le q \le N_2$ is checked as follows:

The result in Fig \ref{normalp} shows all the components are compiled, then components are composited with \textsf{NORMAL} property, and \textsf{NORMALSYSTEM} is not violated the \textsf{NORMAL} property.
\vspace{-0.5cm}
The progress of \textsf{NORMALSYSTEM} is checked:

\vspace{-0.8cm}
 \begin{Verbatim}[fontsize=\footnotesize]
Progress Check...
-- States: 409 Transitions: 870 Memory used: 1633516K
No progress violations detected.
Progress Check in: 6ms
 \end{Verbatim}
The deadlock of \textsf{NORMALSYSTEM} is checked here:

 \begin{Verbatim}[fontsize=\footnotesize]
Analysing...
-- States: 409 Transitions: 870 Memory used: 1544363K
No deadlocks/errors
Analysed using Supertrace in: 23ms
 \end{Verbatim}

\noindent This the result shows no deadlock and progress issue and normal property is hold in this model. The LTSA can also provides the violation traces. The violations case can be found in our website. Furthermore, we make our model open access on Github\footnote{\url{https://github.com/yylonly/LTSA}}, you can download and check the model, and make a contribution by sending a pull request, if needed.

\section{Conclusions and Future Work}
In this paper, a nontrivial case study is presented to demonstrate how LTSA modeling and verifying the real-time system. We show how to specify the structure diagram from UML requirement model, and then generate the start-up design model from the structure diagram. Furthermore, we represent a variation law for the steam rate to prevent the problem of state space explosion. We illustrate how to model the explicit and implicit timer in the components of steam boiler system. For the most important effect of our paper, we show the potential power of integrating UML with the model checking tools in requirement elicitation, system design and verification.


In the future, we consider to integrate LTSA with our code generation tools RMCode \cite{DBLP:journals/corr/YangL16b} to support verifying the requirement model. Furthermore, we consider generate code directly from the verified FSP model. Hopefully, this paper should be useful for in industry and academic worlds.


\section*{Acknowledgment}
This work was supported by the National Natural Science Foundation of China (Grant No. 61562011) and the Macau Science and Technology Development Fund (FDCT) (Grant No. 103/2015/A3). 

\bibliographystyle{ieicetr}
\bibliography{ieice}




 \end{document}